Article

# Enhancing Accuracy and Efficiency in Calibration of Drinking Water Distribution Networks Through Evolutionary Artificial Neural Networks and Expert Systems

## Graphical abstract

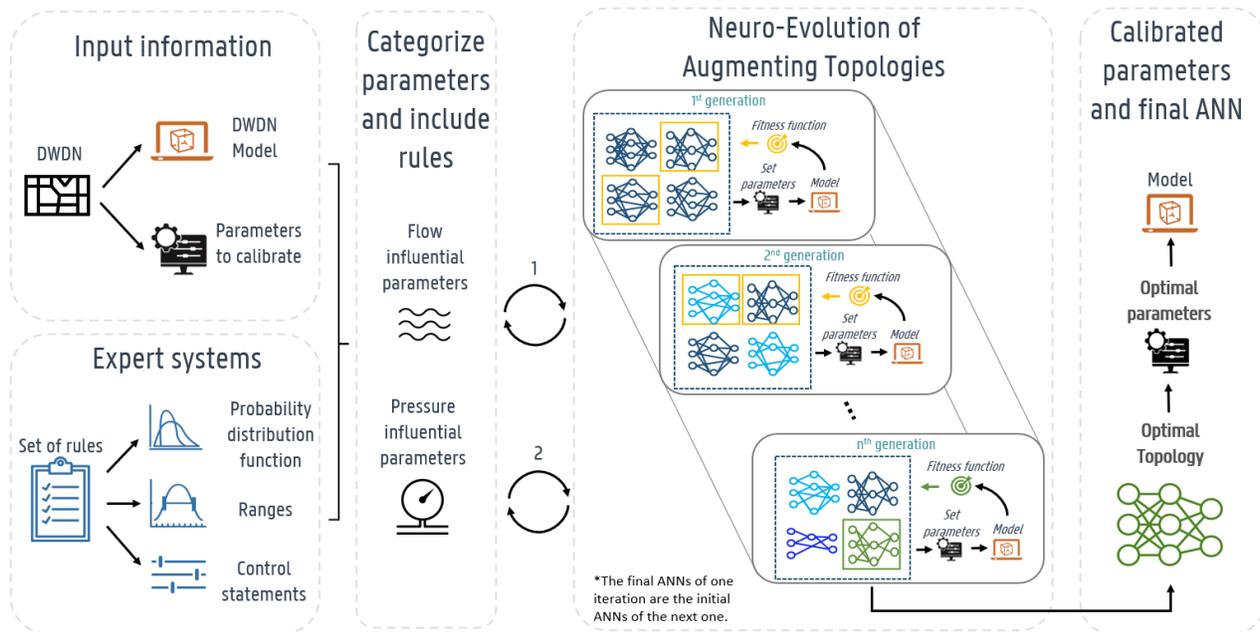

## Authors

Cristian Gomez, Kimberly Solon, Pieter-Jan Haest, Mark Morley, Ingmar Nopens, Elena Torfs

## Correspondence

cristiancamilo.gomezcortes@ugent.be

## Highlights

- A comparison of existing calibration methods for DWDN was performed
- An automated calibration framework for DWDN is proposed based on expert systems and neural networks
- The framework allows the transfer and storage of calibration information between subsequent calibration efforts.
- A good global calibration of parameters was achieved based on limited measurements and limited computation time.
- Expert systems method integrates the case-specific knowledge from stakeholders.

## In brief

A new calibration framework for DWDN capable of transferring calibration information was proposed. Furthermore, the expert systems method integrates the specific knowledge from stakeholders. Finally, global calibration of parameters was achieved with limited measurements available.

Article

# Enhancing Accuracy and Efficiency in Calibration of Drinking Water Distribution Networks Through Evolutionary Artificial Neural Networks and Expert Systems

Cristian Gomez [a,b,*], Kimberly Solon [a,b], Pieter-Jan Haest [d], Mark Morley [e], Ingmar Nopens [a,b], Elena Torfs [a,b,c]

a. BIOMATH, Department of Data Analysis and Mathematical Modelling, Faculty of Bioscience Engineering, Ghent University, Coupure links 653, 9000 Ghent, Belgium

b. Centre for Advanced Process Technology for Urban Resource Recovery (CAPTURE), Frieda Saeysstraat 1, 9000 Ghent, Belgium

c. modelEAU, Département de génie civil et de génie des eaux, Université Laval, 1045, avenue de la Médecine, Québec G1V 0A6, Canada

d. De Watergroep, Rue du Progrès 189, 1030 Brussels, Belgium

e. KWR Water Research Institute, Groningenhaven 7, 3433 PE Nieuwegein, The Nederlands

* Correspondence: cristiancamilo.gomezcortes@ugent.be

## ABSTRACT

The importance of drinking water distribution networks (DWDNs) as critical urban infrastructures has led to the development and utilization of models for the analysis, design, operation, and management of DWDNs, to ensure optimal efficiency and water quality. In order to provide models that accurately represent real-world behavior and characteristics of an actual DWDN, model calibration is an essential and crucial procedure (Alves et al., 2014). However, since DWDNs are generally large, underground networks, data availability for model calibration is often an issue. In this paper, we introduce a novel automatic calibration methodology called Expert Systems and Neuro-Evolution of Augmenting Topologies (ES-NEAT). The proposed methodology leverages the power of Expert Systems (ES) and genetic algorithms for the evolution of neural network topologies to efficiently search for the optimal solution of high dimensional calibration problems while maintaining moderate computational effort. One of the key strengths of ES-NEAT lies in its ability to achieve high accuracy even with limited availability of measurements, addressing the inherent uncertainty in real-world DWDNs. By integrating specific knowledge provided by different stakeholders using the ES methodology, the framework offers a flexible approach that adapts to the unique characteristics of each drinking water distribution network. Moreover, the methodology is designed to store calibration information and transfer it in a structured format for use in subsequent calibration processes, increasing efficiency and ensuring generalizability. The method was successfully applied to a benchmark network model as well as a real-case study of a DWDN in Flanders, Belgium.

# 1. Introduction

Drinking water distribution network (DWDN) models are used for a wide range of purposes. Decision-makers rely on these models for system planning and optimization, performance assessment, cost analysis, and leak detection (Burgschweiger et al., 2009). DWDN models typically contain all the topological and physical information of the elements that make up the network, such as diameters, roughness, minor losses, lengths, and elevations of pipes, nodes, reservoirs, and tanks, among others. As well as consumption information, such as demand and demand patterns at nodes and supply patterns at reservoirs and tanks. From this information, equations are used to determine, among other things, the hydraulic behavior of the network. That is, the flow in the pipes and the pressure at the nodes are calculated in space and time. Both commercial (ex. InfoWorks (Autodesk, 2024)) and open source (ex. EPAnet (Rossman, 2020)) software packages are available to support hydraulic model development for DWDN. However, models must undergo calibration before they can be useful, to establish their accuracy and trust their predictions, or to make rational generalizations out of the model results. A calibration procedure for DWDN models refers to a systematic and rigorous process that involves comparing measurement data coming from sensors, meters, or analytical instruments of parameters such as pressure, or flow rates, against the model outputs with a known level of accuracy. By analyzing discrepancies between measured and modelled values, necessary adjustments can be made to the parameters of the DWDN model, ensuring the correct representation of actual DWDN conditions. An objective function is formulated to assess the discrepancy between the simulated results obtained from the hydraulic model and the observed data collected from the field. Commonly used objective functions include the Root Mean Squared Error (RMSE), the Nash-Sutcliffe Efficiency (NSE), and the Mean Absolute Error (MAE).

The success of the calibration process depends on the parameters selected, the information (data) available and the optimization methodology implemented (Cheng & He, 2010). The most relevant parameters and realistic parameter ranges should be selected to get the most out of the optimization methodologies while prioritizing computational power (Meirelles et al., 2017). Demand patterns, friction factors, leak location/s, and valve positions are some of the most common parameters to determine in the calibration process of hydraulic models of DWDN (Jadhao & Gupta, 2018). However, it is not trivial to determine which are the significant parameters to calibrate, based on the particularities of the DWDN. Moreover, the number of calibration parameters in a DWDN model typically outnumbers the available measurements making the parameter calibration of DWDN an ill-posed problem. Through regular calibration, measurement uncertainties can be minimized, ensuring consistent reliable, and accurate model results. Whereas the calibration process will have to be carried out multiple times over time, the execution time of the calibration process and the resulting model accuracy are significant factors to be considered.

The lack of sufficient and good quality information is a challenge that the calibration methodology must overcome to accurately estimate model parameters. In the case of

DWDNs, measuring flow and pressure in the network on a recurring basis is a costly and difficult task. This is because the monitoring equipment is difficult to install, operate and maintain. Because of this, there is a limited number of measurements along the network, which makes it very difficult to accurately calibrate the DWDN models. Even at present, many DWDN companies are still in the process of designing their flow and pressure monitoring and measurement schemes. Additionally, the presence of errors or noise in the measurements increases the uncertainty in the calibration process (Do et al., 2016). Nevertheless, it is necessary to have as much reliable information as possible from the network. On one hand, pressure measurements are essential to determine the hydraulic behavior of the pipeline sources (i.e., reservoirs, tanks, pumping stations), and subsequently, this will have an impact on the estimation of leakage and energy losses in the DWDN (Berardi and Giustolisi, 2021). Flow measurements, on the other hand, are crucial to maintaining a mass balance in the network, and subsequently, to estimate the different consumption patterns at each node of the network and quantifying the mass loss in DWDN (Sanz & Péreza, 2014).

The calibration process involves the application of optimization techniques to iteratively adjust the model parameters. Initial values are assigned to the model parameters' The initial values serve as starting points for the optimization process and aid in accelerating the convergence of the calibration algorithm. The optimization algorithm is iterated multiple times, with each iteration adjusting the model parameters based on the performance of the previous iteration. The calibration process continues until an acceptable level of convergence is reached, indicating a satisfactory match between the model and the real-world data.

Various approaches have been proposed as an optimization methodology to use within the calibration process. Manual calibration, mathematical and gradient-based optimization, and evolutionary algorithms are the most popular techniques to solve the calibration problem. Due to the particularities of each DWDN, each methodology has its advantages and disadvantages, and there is no definitive solution to the calibration problem.

Manual calibration based on expert knowledge presents an affordable and simplistic method for tackling calibration. This approach allows flexibility to adapt to the particularities of each network. However, it can be time-consuming when dealing with many parameters or complex systems. It also lacks the precision and accuracy of automated calibration procedures and requires high expert knowledge (Berardi & Giustolisi, 2021).

Among the mathematical and gradient-based optimization methodologies, singular value decomposition (SVD) presents promising results for the calibration process (Cheng and He, 2010). The SVD algorithm can handle an arbitrary number of parameters when the solution to the equations is overdetermined, even-determined, and underdetermined. Nevertheless, prior information is required in some cases to control the variability of the parameters and achieve better convergence of the results (Sanz and Péreza, 2014). Furthermore, nonlinear programming (NLP) is another recommended methodology able to tackle the calibration problem (Jadhao & Gupta, 2018). The nonlinearity approach of the methodology allows to better identify relationships and limits within the search space. However, it also introduces an increased computational complexity and potentially requires more sophisticated algorithms and mathematical techniques which may require additional expertise and resources for implementation and solving. Finally, the enhanced global gradient algorithm (GGA) is a methodology capable of efficiently searching for the global optimum by intelligently exploring the search space through the use of global and local search strategies (Giustolisi & Berardi, 2011). It also takes advantage of the averaging effect of uncertainties surrounding boundary conditions in a DWDN. Still, the algorithm suffers from slow convergence in certain cases and requires a lot of parameter tuning, which can be time-consuming and require expertise.

Lastly, examples of implemented evolutionary algorithms, inspired by natural selection, present a variety of useful methodologies. First, the genetic algorithm is a reliable methodology that explores different possibilities in the search space with an evolutionary approach (Nicolini et al., 2010). Genetic algorithms can adapt to changing environments and a large number of parameters. However, the algorithm is computationally expensive and can take a lot of time to converge to a satisfactory answer (Do et al., 2016). Secondly, ant colony optimization allows for robust exploration of the solution space. Also, it is capable of finding near-optimal solutions quickly and can adapt to changing problem conditions. However, it can be computationally expensive and can get trapped in a local optimum and struggle with dynamic or highly nonlinear problems (Dini & Tabesh, 2014). Thirdly, the imperialist competitive algorithm overcomes local optima and facilitates global optimization. Nevertheless, it requires careful parameter tuning and can produce high computational complexity (Moasheri & Jalili-Ghazizadeh, 2020). Finally, a blended methodology of artificial neural network (ANN) and particle swarm optimization was explored by Meirelles et al. (2017). Initial implementation of ANN allows for a reduction in the degrees of freedom of the problem, which improves the performance of the particle swarm technique. However, the parameter tuning complexity highly increases and can be propagated through the methodology (Meirelles et al., 2017).

Most of the methodologies described above present similar drawbacks to the general calibration problem. In summary, a large number of parameters are necessary for a successful calibration process. This makes the computational effort in many cases large and may result in non-convergence of the solution. Additionally, a limited number of measurements are available causing the calibration problem to present a complex search space with multiple local optima and steep gradients that make exploration difficult, increasing the complexity of the system.

Finally, it is not explicitly mentioned in any methodology, but there is no feasible way to transfer the knowledge from one calibration effort to the next. Consequently, it is necessary to restart the calibration process again when required (for example whenever the model performance is no longer satisfactory or whenever significant changes in the design, operation, or general state of the network occur). A calibration method that is able to transfer in a structured way the knowledge acquired from one calibration process to the next would significantly speed up subsequent calibration efforts. This involves transferring the interaction between the parameters, their best-fit ranges, and their most probable values according to the model outputs. During manual calibration, this process is carried out consciously or unconsciously by the experts as they gain experience with the models and allowing them to perform the calibration process faster when it is required again. However, today it is a step that most optimization algorithms omit due to the lack of structure to perform this task.

This study aims to: (i) Define a calibration framework for the global calibration of the large number of parameters involved in DWDN models, with moderate computational expenditure and high accuracy for the limited measurements available; (ii) Integrate the specific knowledge provided by the different stakeholders into the calibration framework in a reproducible way to boost its accuracy; and (iii) Store information related to the calibration process and transfer it in a structured way to the next calibration process, thus reducing the subsequent calibration efforts.

Therefore, towards the development of an automatic calibration framework, this paper first explores and compares the performance of several existing methodologies in the task of calibrating a benchmark network. Subsequently, the methodologies with the best performance were selected, taking into account the objectives set. As a result, the ES-NEAT automatic calibration methodology is created, which incorporates evolutionary topologies of artificial neural networks and expert systems. Finally, the performance of the ES-NEAT methodology is tested for a real DWDN case.

## 2. Material and Methods

### 2.1. *Case-studies*

#### 2.1.1. *Benchmark network*

As a first step, different existing optimization methods from the literature are compared, and their performance is analyzed. For this, a medium-sized benchmark was taken as a starting point to ensure comparison under controlled conditions. By assessing factors such as the convergence speed, accuracy of calibration, and the capability to transfer calibration information to subsequent calibration processes, valuable insights into the strengths and limitations of each method are gained. The comparative analysis enables to identify the most effective or promising optimization approach.

The chosen benchmark is the Fossolo network (Wang et al., 2015), henceforth referred to as the benchmark network, the layout of which is shown in Figure 1. The benchmark network consists of 58 pipes, 36 demand nodes, and a reservoir with a constant head of 121.00 m. Polyethylene is the chosen material for all the pipes, a roughness coefficient of 0.0015 mm is uniformly assigned to them. It is ensured that the minimum pressure head at all the demand nodes remains at 40 m. 3 nodes and 3 pipes were selected to represent the elements measured in the benchmark network, their distribution can be seen in Figure 1.

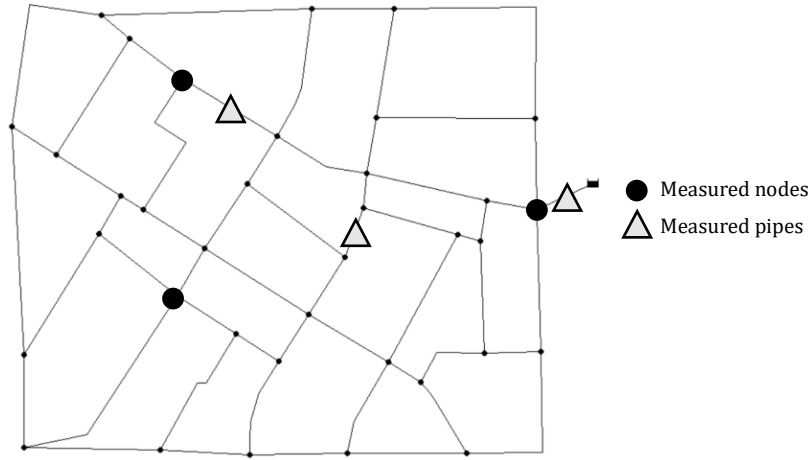

Figure 1. Benchmark network layout

### *2.1.2. Real drinking water network*

To evaluate the performance of the new proposed calibration framework, a real network was used. The actual network located in Belgium, shown in Figure 2 (a), includes 4158 pipelines, 5197 demand nodes, and 5 reservoirs with fixed heads within 1 and 5 bars. The distribution and elevation of the network are between 35 and 100 m, with an average reservoir head of 106 m. The pipe materials are mainly high-density polyethylene, PVC, concrete, cast iron, and steel. The locations of the measured nodes and pipes are shown in Figure 2 (b) for flow measurements and Figure 2 (c) for pressure measurements, respectively. A total of 42 nodes and 8 pipes were measured in this network, in which 3 nodes and 2 pipes were used for validation. One year of measurements with a frequency of 15 minutes were supplied were an average week of consumption was created for this study. This network contains all the information available from the water distributor, regarding the topology, and physical characteristics of the elements and hydraulic behavior. As well, a first manual calibration was performed on the parameters. The manual calibration is the one that is going to be compared against the methodology developed.

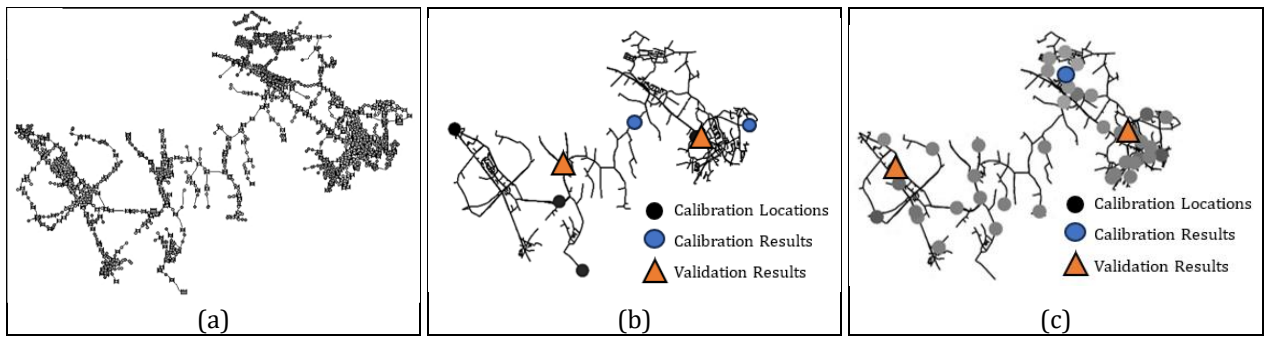

Figure 2. (a) The layout of the real network (a) and locations of the flow (b) and pressure (c) measurements

## 2.2. *Optimization methods*

A list of calibration methods that were tested is provided in Table 1. This list is not intended to be exhaustive but rather takes some methodologies from the most popular groups of optimization methods, such as mathematical and gradient-based optimization, evolutionary algorithms, and artificial neural networks, which have also been used in DWDN models in other studies. For a more extensive list, the work developed in the Ostrich automatic calibration tool software compiles multiple methodologies for calibration problems (Shahed Behrouz et al., 2020).

Table 1. List of optimization methods tested and references of their use during a calibration procedure of DWDN systems.

| Optimization methodology | Abbreviation | Reference |
|---|---|---|
| Monte Carlo | MC | (Houska et al., 2015) |
| Markov-Chain Monte-Carlo | MCMC | |
| Maximum Likelihood Estimation | MLE | |
| Latin-Hypercube Sampling | LHS | |
| Simulated Annealing | SA | |
| Shuffled Complex Evolution Algorithm | SCE-UA | |
| Differential Evolution Adaptive Metropolis Algorithm | DE-MCz | |
| Robust Parameter Estimation | ROPE | |
| Artificial Bee Colony | ABC | |
| Differential Evolution Adaptive Metropolis | DREAM | |
| Fitness Scaled Chaotic Artificial Bee Colony | FSCABC | |
| Fourier Amplitude Sensitivity Test | FAST | |
| Genetic Algorithms | GA | (Do et al., 2016) |
| Particle Swarm Optimization | PSO | (Haidar et al., 2021) |
| Levenberg-Marquardt Algorithm | LMA | (Moré, 1978) |
| Harmony Search | HS | (Li et al., 2020) |
| Imperialist Competitive Algorithm | IC | (Hosseini & Al Khaled, 2014) |
| Artificial Neural Networks | ANN | (Stanley & Miikkulainen, 2002) |
| Neuro-Evolution of Augmenting Topologies | NEAT | |

Throughout the analysis, the calibration methodologies are scrutinized based on several performance criteria. First and foremost, their ability to achieve the most relevant calibration objectives is assessed, which includes obtaining accurate estimates for the large number of variables involved in the DWDN model. The calibration accuracy is evaluated with the RMSE. A lower RMSE value indicates that the model fits the data well and has more precise predictions, while higher values suggest more error and less precise predictions. (Jain and Singh, 2003).

Additionally, the importance of moderate computational performance is emphasized, as DWDNs often consist of complex networks with a considerable number of parameters, requiring efficient optimization techniques. For this criterion, the convergence speed was taken as a performance measure. Furthermore, how well each methodology enables the transfer of calibration information to subsequent calibration processes is investigated, streamlining future model-tuning efforts.

Ultimately, the analysis presented in this section will serve as a guide for selecting the most appropriate calibration methodology, one that optimally aligns with the specific objectives and constraints of DWDNs.

## 2.3. *Novel calibration methodology: ES-NEAT*

Throughout this section, the structure and operation of the ES-NEAT automatic calibration methodology will be presented. This methodology was created as a result of the comparative analysis carried out with the optimization methodologies, presented in section 2.2. and 3.1, and the calibration objectives presented in section 1. The methodology represents a reliable alternative for the global calibration of highly parameterized systems, as well as the possibility of storing and transferring calibration information. The conceptual scheme of this methodology can be seen in Figure 3.

The algorithm initiates by capturing the topological structure of the water distribution network and the associated physical attributes of its constituent elements, including pipes and junctions. From this information, an EPANET model is formulated, encompassing essential properties such as network topology, pipe diameter, surface roughness, minor losses, demand base, and consumption patterns. Subsequently, the calibration process selectively targets unknown parameters crucial to the system's hydraulic behavior. Specifically, the parameters subjected to calibration in this paper are the following: pipe diameter, pipe roughness, minor losses in pipes, nodal base demand, leaks in the system, and the influence of valve-induced minor losses.

In the calibration procedure, the global objective is the optimization of flow and pressure along the network. However, the challenge arises from the potential interdependence among parameters, wherein adjustments to enhance one aspect may inadvertently introduce compensatory effects that impact the other, either through magnification or diminishment beyond their inherent values. To circumvent this issue, a segregation of parameters into two distinct categories is proposed: those exerting a more pronounced influence on flow and those exhibiting a greater influence on pressure. The rationale behind this categorization lies in mitigating the potential for inter-parameter compensation during the iterative calibration process.

The methodology adopted entails a sequential calibration approach wherein parameters influencing flow dynamics are prioritized for adjustment first since flow is more sensitive to variations than pressure. After obtaining an adequate calibration for the flow, proceed to calibrate the parameters related to the pressure dynamics. This sequential refinement strategy continues iteratively until convergence is achieved. By adhering to this iterative sequence, the likelihood of parameters compensating for one another is substantially reduced, thereby fostering a more robust calibration outcome.

The proposed methodology relies on the integration of expert systems with NeuroEvolution of Augmenting Topologies (NEAT) to construct neural networks tailored for calibration purposes. The methodology was developed based on the work done by Gomez et. al. (2021). This innovative technique leverages the adaptability and problem-solving capabilities inherent in neural networks, mimicking the intricate connectivity patterns observed in biological brains, known as network topology. This topology contains the connections between relevant variables in the calibration process, crucial for determining optimal parameter configurations.

The challenge arises in establishing an appropriate topology since determining the proper topology is not a trivial problem and usually requires a calibration process in itself. NEAT, an evolutionary algorithm for augmenting topologies, addresses this by iteratively evolving neural network topologies. Analogous to genetic algorithms, NEAT commences with diverse randomly generated topologies in the initial generation. The network inputs correspond to the parameters and characteristics of the network that are known and influence the calibration variables, such as pipe lengths, network connectivity, or valve type, among others. These topologies undergo evaluation against a predefined objective function, comparing their outputs with empirical data to identify superior performers. In this study, the methodology compares the flow and pressure curves produced by the EPANET model with the real measured flow and pressure curves in some elements in the DWDN. Selected topologies then propagate to subsequent generations, iteratively refined through evaluation and selection, culminating in a topology best suited to derive optimal calibrated parameters.

NEAT, initially developed for robotic control and autonomous learning endeavors, was conceptualized as an optimization methodology rather than a calibration protocol. In the domain of autonomous learning, NEAT has demonstrated robust performance, particularly in systems characterized by high-dimensional parameter spaces (Mnih et al., 2013). The innovation in applying NEAT to calibration contexts lies in its capacity to efficiently navigate expansive parameter landscapes, mitigating the risk of local optima while dynamically adjusting network size. Given the absence of direct measurements for selected parameters in DWDN calibration, NEAT presents an effective means of globally calibrating parameters in highly parameterized systems, aligning with observed flow and pressure profiles acquired in the field.

Preceding the initiation of NEAT, a prerequisite involves the explicit definition of hyperparameters governing the methodology. These hyperparameters encompass the fitness criterion and threshold, which intricately define the optimization direction (maximization, minimization, etc.), along with the stop criteria. The maximum number of generations represents a vital parameter acting as the termination criterion for maximum iterations when the simulation fails to achieve the defined fitness threshold within the designated iteration limit. The population size defines the number of ANNs of each generation that are going to be generated and tested.

Several other critical hyperparameters shape the operational characteristics of NEAT. The activation function for nodes, a fundamental attribute in ANN, defaults to the Sigmoid function. The number of inputs and outputs is determined by the nature of the problem, and the configuration of hidden nodes within these layers is specified in the initial iteration. The initial connection state dictates the network's foundational connections, defaulting to a fully connected state.

Moreover, the rate of addition or removal of nodes and connections assumes significance, influencing the frequency at which connections or nodes are incorporated or eliminated from the network, thereby contributing to the pursuit of optimal network topology. These aforementioned hyperparameters collectively constitute the essential parameters requisite for the proper execution of the NEAT algorithm.

Furthermore, the method yields a neural network that encapsulates all calibration-related information, including parameter behavior, inter-parameter interactions, and pertinent parameter ranges, inherently encoded within its topology. Through the application of transfer learning, this repository of knowledge can be leveraged in subsequent calibration efforts. Transfer learning, in this context, denotes the process whereby an algorithm assimilates knowledge from one problem domain to another (Tan et al., 2018). Consequently, future calibration benefits from accelerated convergence, with reduced overhead associated with artificial neural network optimization, necessitating a reduced investment of time and expertise. Such adaptability not only mitigates the risk of overfitting but also augments algorithmic efficacy over successive iterations.

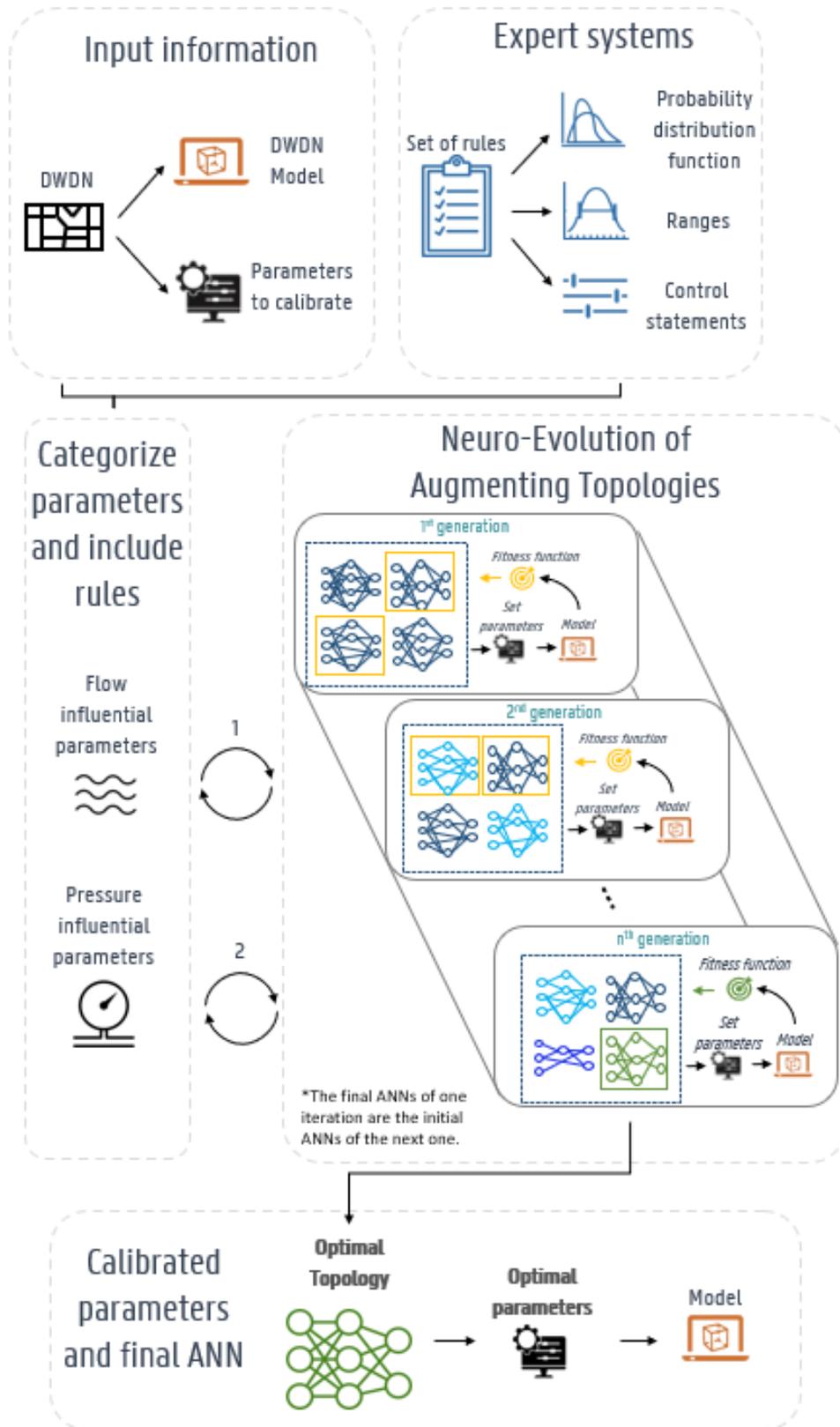

Figure 3. ES-NEAT methodology

While NEAT governs the internal structure and outputs of the methodology, the input parameters are informed by expert systems. These systems encapsulate domain knowledge, encompassing parameter characteristics such as ranges, probability distributions, and control statements derived from literature or empirical studies. Moreover, recognizing the significance of domain-specific knowledge and expertise, the calibration framework incorporates the specific knowledge provided by different stakeholders. The experience and knowledge about a DWDN are crucial to limit the search space of the parameters to calibrate. Each DWDN possesses unique characteristics and operational requirements, necessitating a flexible approach to calibration. Through an expert systems methodology, stakeholder knowledge is integrated into the calibration process, enabling the framework to adapt to the particularities of each DWDN. This holistic approach enhances the model's representation of the network's behavior, leading to more accurate and reliable results.

To incorporate specific knowledge into the calibration framework, an expert system (ES) methodology was implemented (Kaisler, 1986). The ES methodology systematically develops computer-based systems that emulate human expertise in specific domains. Utilizing knowledge engineering techniques, it captures and formalizes the reasoning processes of domain experts. The ES comprises a knowledge base with explicit rules, facts, and heuristics guiding decision-making and problem-solving capabilities through an inference engine. Expert systems play a valuable role in DWDN calibration by incorporating domain-specific knowledge and using logical reasoning to generate constraints and recommendations for improving calibration.

The ES can include rules provided by water network experts, capturing insights on the relationships between system parameters, hydraulic behavior, and calibration criteria. By employing logical reasoning and inference mechanisms, the ES processes available data and knowledge, generating customized calibration ranges for each DWDN model parameter based on its unique characteristics. This tailored approach enhances the calibration process, providing decision support for adjusting parameters. These parameters include base demands, leaks, friction factors, minor losses, and valves. The diameter can be a variable to be taken into account, however, it was not considered in the calibration of this study because its value was known.

Within the calibration procedure of parameters associated with flow, the following information was taken into account in formulating the rules. Regarding base demands, the implemented rules consider factors such as population growth, seasonal variations, and distinctive demand patterns stemming from industrial and domestic sectors. Additionally, the calibration constraints applied to leakages, involved specifications regarding leak locations, magnitudes, and variations in water loss by sectors.

Conversely, in the calibration process on parameters associated with pressure, information regarding roughness and minor losses was taken into account in formulating the rules. In the context of roughness, integrated considerations for pipe materials and addressed erosive effects influenced by the age of the pipe. Specific types of pipe fittings were accounted for in the calibration of minor losses along the pipeline. Lastly, based on the specific valve type and operational status, the minor losses of the valves were assigned.

Figure 4 illustrates the considerations taken into account for defining the calibration ranges, demonstrating the ES's ability to adapt to individual parameter characteristics, thereby optimizing the calibration process.

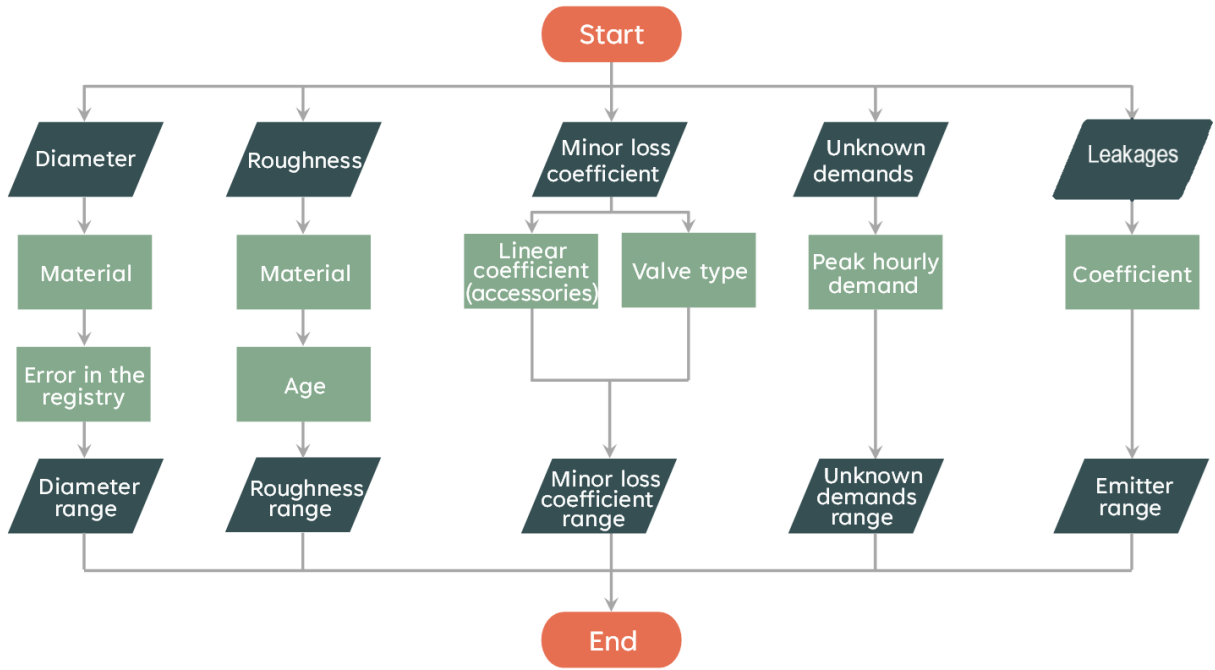

Figure 4. Rules in the expert system (ES) for a DWDN

## 2.4. *Model implementation and simulation*

Hydraulic simulation is performed to model the fluid behavior in the DWDN, analyzing aspects such as fluid flow, pressure, and interactions with other components under various conditions. In this study, the open-source software application EPANET (Rossman, 2020) was used.

The different optimization methods were implemented in the Python programming language, due to its flexibility and the libraries available in it. Water Network Tool for Resilience (WNTR) version 1.0.0 was implemented to connect EPANET version 2.2 and Python 3.10.9 (Klise et al., 2017). The WNTR package is a Python library built upon EPANET that provides an interface for working with EPANET models in a more accessible and flexible manner.

The implementation of different calibration methodologies was carried out with the help of specialized Python libraries such as SPOTPY (Statistical Parameter Optimization Tool for Python) version 1.6.2, NEAT (Neuro-Evolution of Augmenting Topologies) version 0.92, TensorFlow (Abadi et al., 2016) version 2.10, etc.

The experimental computations were performed on a computer system with the following hardware specifications: the processor is an 11th Gen Intel(R) Core(TM) i5-1145G7, operating at a base frequency of 2.60 GHz with a maximum turbo frequency of 4.40 GHz. The system featured 16.0 GB of installed RAM, with 15.7 GB available for use. The computer operated on a 64-bit system, running on Windows 10 operating system.

## 3. Results and discussion

### 3.1. *Comparison of existing calibration methods on a benchmark network*

The various optimization methods listed in Section 2.2 were run on the benchmark network. To calibrate the benchmark network model, the roughness and minor losses parameters of the pipes, as well as the base demands and leakage coefficients in the nodes, were modified. Root mean square error (RMSE) was used to measure the difference between the model results and the actual measured values for flow and pressure. An RMSE of 19.81 was obtained for the pre-calibrated network. This value is the basis for the comparison of the various optimization methodologies tested. Figure 5 shows the performance of each methodology. The comparative analysis compares the minimum values obtained throughout the calibration process among the analyzed methodologies.

In pursuit of an expeditious calibration methodology, a predefined threshold of 1000 iterations was established as the criterion for assessing the viability of candidate algorithms within the calibration process. This numerical threshold serves as a basis for determining the acceptability or rejection of candidates for inclusion in the final methodology. The assessment considers the candidate's ability to attain a satisfactory level within the specified simulation limit. Additionally, throughout these 1000 iterations, an examination is conducted of the speed convergence with which a candidate algorithm achieves a value below the acceptable threshold of the objective function. The overall speed convergence was taken as an overall measure since the methodologies evaluate the objective function at different rates.

This comprehensive evaluation framework considers not only the attainment of desirable performance levels but also the convergence speed of each methodology. The acceptance criterion is further refined by stipulating a threshold value of 3 for the RMSE function. This formalized approach ensures that the chosen methodology not only demonstrates proficiency in reaching acceptable performance levels but also exhibits prompt convergence and stability, aligning with the overarching objective of expeditious and reliable calibration.

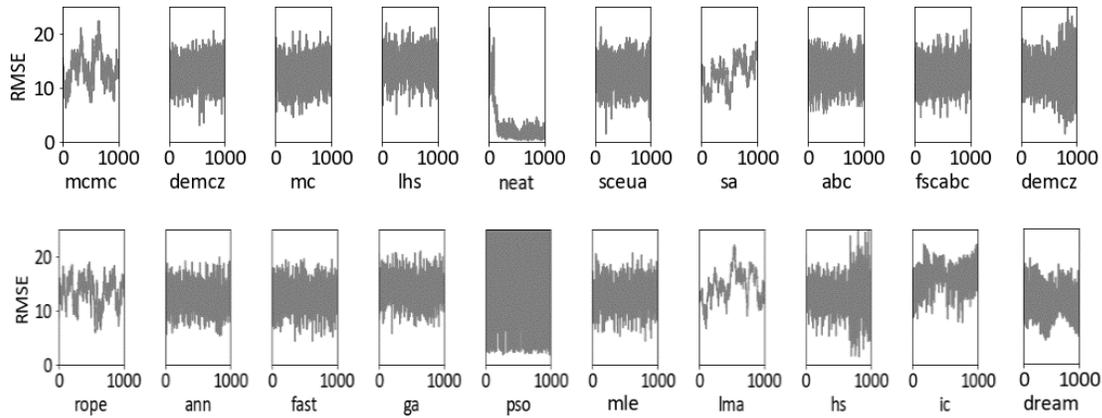

Figure 5. Performance of the various optimization methodologies in the calibration of the benchmark network in terms of RMSE

Among the calibration methodologies, NEAT, SCEUA, and PSO algorithms performed best with final RMSEs of 1.85, 2.37, and 2.62, respectively. However, recalling the objectives previously described for this study, NEAT was chosen as the best optimization methodology not only because it gave the lowest average RMSE but also because of its ability to store the information in a neural network that can be later recovered. Furthermore, the algorithm also showed a more stable evolution of the fitness function over time. The selected methodology will be implemented with the available information. From this information, a part will be kept for the validation process, where the calibration result will be evaluated at locations where the mass and pressure curves were not adjusted by the calibration algorithm. The objective is to reduce the overfitting of the algorithm at the limited measurement points available. Ensuring a better overall performance of the calibration process in the whole network.

## 3.2. *Performance of the ES-NEAT calibration framework*

In this study, the effectiveness of the novel framework ES-NEAT was tested on a real DWDN shown in Figure 2. The calibration process of about 20,000 parameters achieved an RMSE of 0.33, calculated considering both flow and pressure measurements. RMSE values are not comparable between case studies because they depend on the amount and magnitude of data. However, as previously mentioned, values closer to zero represent a better fit. As a general rule for this study, a value of less than 0.6 is considered good. This value is accepted by the community as a good measure of fit. Because of this, the value obtained indicates a reasonably good accurate representation of the observed data.

The results were achieved in 12 hours with a total of 10000 DWDN simulations. 100 generations with 100 individuals each were applied to achieve these results. This is a low amount of simulations considering that the number of parameters to calibrate is around 20,000. The first 10 generations (1000 simulations) take about 20% of the total calibration time because the algorithm takes an explorative approach before focusing on exploitative improvement. Finally, The ANNs used as input parameters the pipe lengths, network connectivity, pipe diameters, and node elevations, among others.

The hyperparameters implemented were minimization as the fitness criterion and a threshold of 0.1 for the RMSE. A total number of generations (iterations) of 100, with a population size of 100. A clamped and sigmoid activation function were used in the network. More information on these activation functions can be found in the NEAT documentation. The initial connection state was set as fully connected, with no hidden nodes. The addition rate connections and nodes were augmented to 0.7 and 0.4 respectively and a removal rate of 0.4 and 0.2 respectively.

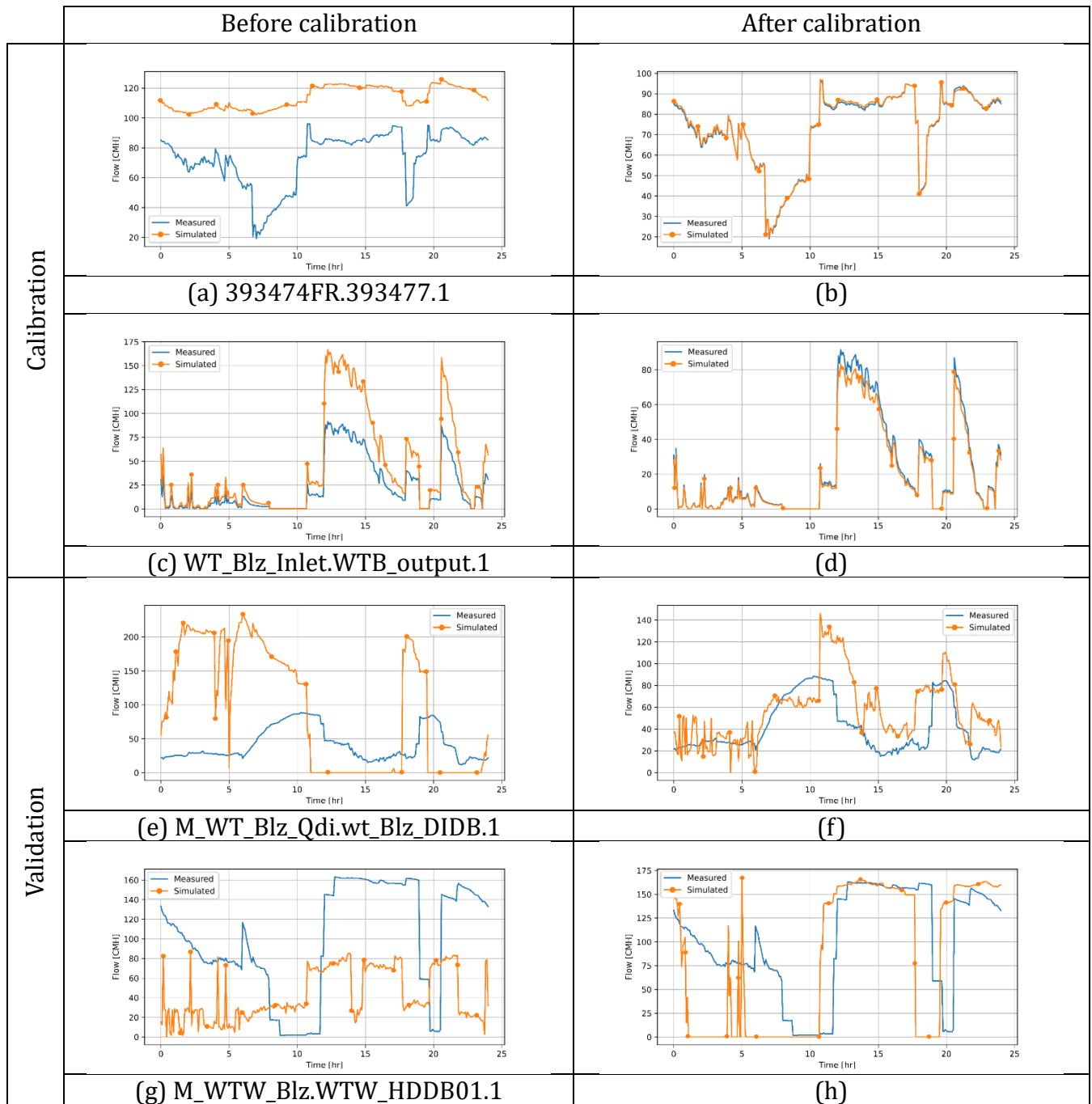

Figure 6. Model simulation results for flow before and after calibration

Figures 6 (a) and (c) represent the flow calibration results before implementing the ES-NEAT framework, respectively. The results in these graphs present the manual calibration carried out by the water distributor, which represented a time-consuming and costly task, due to the lack of implementation of an optimization methodology in the calibration process. Before implementing the novel global calibration framework, there are evident discrepancies between the simulated flow values and the observed data in the measured pipes. However, Figures 6 (b) and (d) present the flow calibration results after implementing the ES-NEAT framework. After calibration the lines representing the calibrated flow values align remarkably well with the observed data, indicating a significant improvement in accuracy. The calibration process successfully minimizes the discrepancies, leading to an adequate calibration for flow in the measured pipes.

Figures 6 (e), (f), (g), and (h) show the results of the validation process of the flow measurements. The calibrated model continues to closely match the observed data during the validation phase. The model's ability to replicate the observed flow patterns even during validation demonstrates the reliability and robustness of the calibration achieved using the ES-NEAT framework. Overall, the flow calibration process proves to be successful, as evidenced by the accurate representation of flow in both the calibration and validation stages.

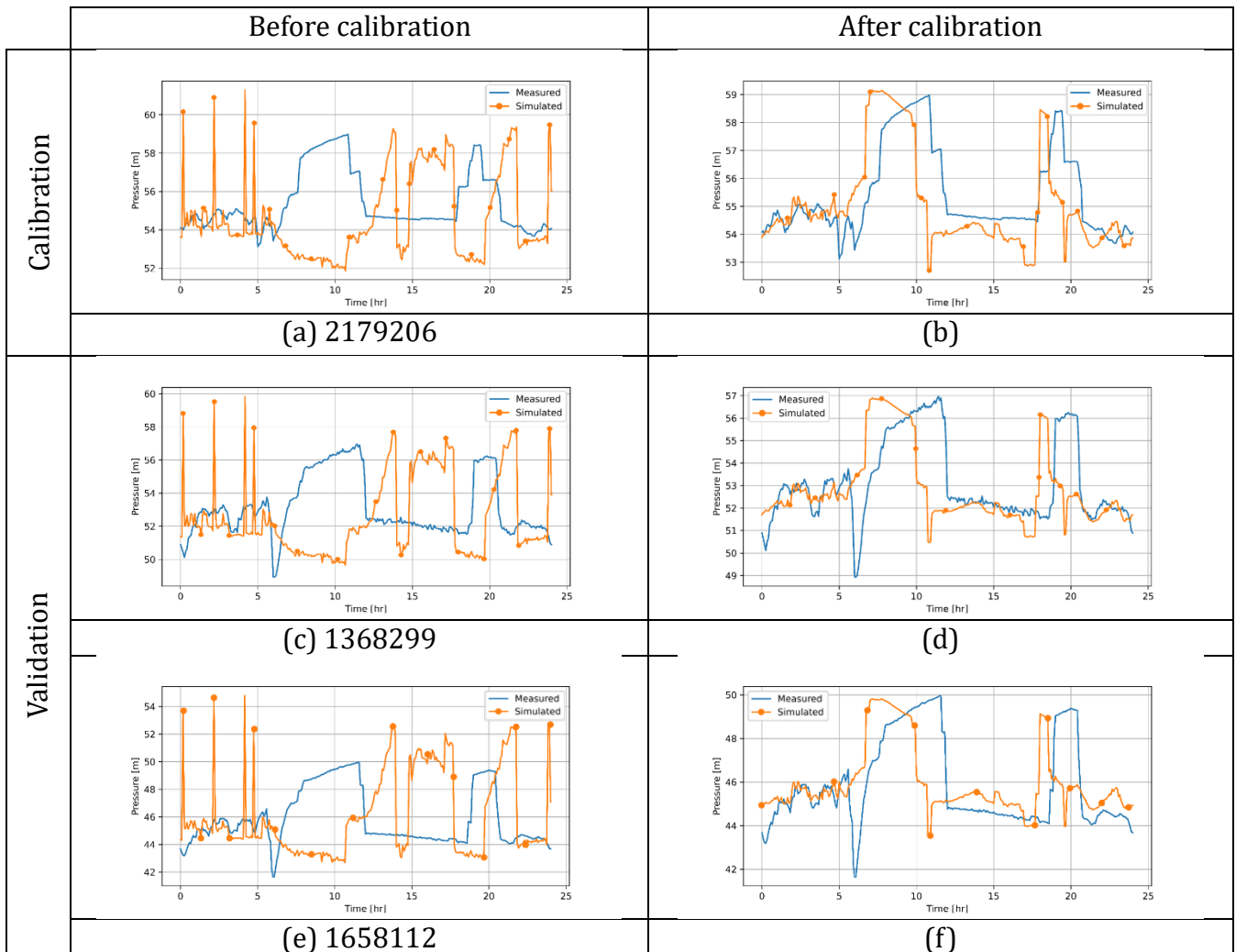

Figure 7. Model simulation results for pressure before and after calibration

In Figures 7 (a), and (b) the pressure calibration results before and after employing the ES-NEAT framework are displayed. Similar to the flow calibration, Figures (a), and (b) indicate discrepancies between the simulated pressure values and the observed data at the measured nodes before calibration. Following the calibration process in Figures 7 (a), and (b), the lines representing the calibrated pressure values demonstrate a remarkable improvement, closely matching the observed data. This significant reduction in discrepancies signifies the framework's capability to achieve an adequate calibration for pressure at the measured nodes.

Figures 7 (c), (d), (e), and (f) present the pressure results for the validation process. During this stage, the calibrated model continues to exhibit close agreement with the observed data, reaffirming the framework's effectiveness in accurately predicting pressure values. The successful calibration of pressure for both calibration and validation stages demonstrates the reliable performance of the ES-NEAT framework in capturing the complex dynamics of the water distribution network. The overall validation phase yielded an RMSE of 0.56, further substantiating the framework's capability to generalize and predict water distribution system behavior, indicating a satisfactory agreement between the observed and simulated values.

The calibration results obtained using the ES-NEAT framework provide strong evidence of its efficacy in optimizing flow and pressure parameters for the water distribution network. The before and after calibration graphs showcase the substantial improvement achieved in accurately simulating the observed data. The framework successfully minimizes discrepancies, leading to an adequate calibration for both flow in the measured pipes and pressure at the measured nodes.

By representing the most important peaks of flow and pressure as well as the overall trend of flow and pressure patterns, ES-NEAT proves its capability to replicate critical aspects of the system's behavior. This level of accuracy is particularly impressive given the limited measurements available, demonstrating the framework's robustness in handling uncertainties and variations in real-world settings.

The integration of specific knowledge from different stakeholders using the expert systems methodology contributes to the framework's adaptability to various drinking water distribution networks. This flexibility allows ES-NEAT to cater to the unique characteristics of individual systems, enhancing its practical utility in diverse operational scenarios.

To position this work among the current state of the art, the results of the calibration results of the novel automatic calibration framework, ES-NEAT, were compared with two previous approaches. The first one implements Particle Swarm Optimization and ANN to calibrate (Meirelles et al., 2017). This paper may have reported similar RMSE values for calibration and validation. However, it is essential to consider the complexity and scale of the networks used in each study. Artificial Neural Networks, by itself, require substantial data and computational resources to train effectively, through backpropagation, which might limit their practicality for large-scale and complex networks. In contrast, ES-NEAT demonstrates a competitive level of accuracy while avoiding the backpropagation process, maintaining moderate computational performance, and making it a more viable option for real-world water distribution networks.

The second approach implements genetic algorithms in the calibration process (Nicolini et al., 2010). Regarding calibration results, both ES-NEAT and genetic algorithms may achieve comparable RMSE values for flow and pressure calibration, indicating that both approaches successfully replicate observed data. However, ES-NEAT's strength lies in its ability to provide a structured and flexible calibration framework that efficiently handles a large number of parameters.

## 3.3. *Transfer learning*

A critical advantage of the "ES-NEAT" framework is its capacity to store all calibration information in an artificial neural network (ANN). An example of the artificial neural network created during the optimization process is presented in Figure 8, where the upper squares represent the input variables, the network inputs correspond to the parameters and characteristics of the network that are known and influence the calibration variables, such as pipe lengths, network connectivity or valve type, among others. The middle circles are the hidden network capabilities. Finally, the lower circles are the output variables, which in the calibration of the DWDN context correspond to pipe diameter, pipe roughness, minor losses in pipes, nodal base demand, leaks in the system, and the influence of valve-induced minor losses, among others. The lines between these elements represent the connections, the darker the color of these lines, the stronger the interaction between the elements. Being a fully connected approach, it ensures that all interactions between variables are contemplated. However, during the evolutionary process, new elements are eliminated or created, maximizing the efficiency of the neural network. This represents a section of the complete neural network, which contains the interactions between the calibrated parameters.

The strength of the connection between two parameters directly influences the magnitude of their connection weights within the neural network architecture. Parameters exhibiting a robust correlation manifest higher connection weights, whereas those demonstrating a weaker relationship are characterized by lower weights or may lack a connection altogether. Consequently, parameters with a closer association exert a more pronounced influence on the calibration process, thereby significantly impacting network behavior.

Additionally, the output layer's response is intricately tied to parameter interactions, with variations in one parameter potentially eliciting corresponding changes in others. This phenomenon underscores the interconnected nature of parameter influence within the calibration framework, wherein alterations in one parameter can reverberate across the network, shaping the collective output response. The ability to use a transfer learning approach in the ES-NEAT method represents an important advantage over other methodologies.

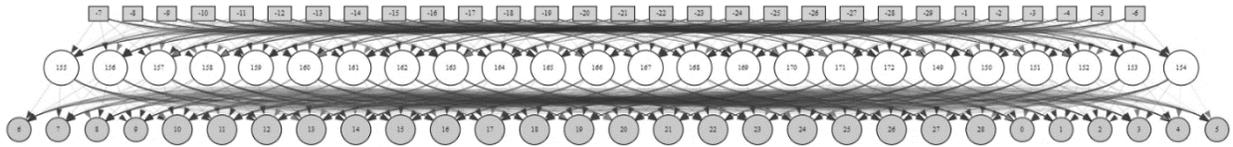

Fig. 1. Section of an artificial neural network embedded with calibration information.

This feature enables researchers and practitioners to access and utilize the calibration data for subsequent calibration processes. The methodology is able to recall and start the next calibration process implementing the previous optimal ANN as a baseline to fast-track the calibration process. The structured format ensures that valuable knowledge gained during previous calibrations is not lost and can be leveraged for further improvements or future network modifications. During the calibration process, the interactions within the parameters prevail and strengthen as calibration accuracy improves, allowing calibration information to be stored within ANN connections.

By implementing the final neural network as the basis of the first generation of the NEAT evolutionary process in the next calibration, a performance improvement in the calibration process can be obtained, which improves as the network is further calibrated (a process that as previously mentioned, is carried out multiple times throughout the useful life of a DWDN). Further studies on the uncertainty of ANNs and their topology should be carried out to limit the search space and optimize the transfer learning process.

## 4. Conclusion

In the presented study, a novel automatic calibration framework for DWDNs was developed, and the methodology was tested in a real network. The main research findings are:

- The novel automatic calibration framework for DWDNs, based on the joint methodology of Expert Systems (ES) and Neuro-Evolution of Augmenting Topologies (NEAT) was described.
- The specific knowledge provided by the different stakeholders was integrated through the ES methodology, which provides a flexible approach to the particularities of each DWDN.
- The calibration framework, developed through assessment of multiple optimization methodologies, presents the best approach for the calibration of a large number of parameters, with moderate computational performance, and high accuracy for the limited measurements available.
- The calibration framework is capable of storing the calibration information and transferring it to the following calibration process with a structured format.

**Declaration of Competing Interest**

The authors declare that they have no known competing financial interests or personal relationships that could have appeared to influence the work reported in this paper.

**Acknowledgments**

This work was developed in collaboration with the Flemish drinking water company De Watergroep and the Dutch research institute KWR through the research project: "Automatic Hydraulic Calibration of Drinking Water Networks".